\documentclass[floatfix,showpacs,preprintnumbers,amsmath,amssymb,aps,twocolumn,superscriptaddress,prl,10pt]{revtex4}

\usepackage{amsfonts}
\usepackage[pdftex]{graphicx}
\usepackage{color}
\usepackage{bm}
\usepackage{multirow}
\newcommand{\mat}[1]{\mathsf{#1}}
\newcommand{\figref}[1]{Fig.~\ref{#1}}
\newcommand{\e}[1]{\text{e}^{#1}}
\newcommand{\cmplxi}{\text{i}}

\renewcommand{\vec}[1]{\mathbf{#1}}
\newcommand{\punc}[1]{\,#1}
\newcommand{\neweqnline}{\nonumber\\}

\newcommand{\eqnref}[1]{Eqn.~(\ref{#1})}
\newcommand{\vecgrk}[1]{\boldsymbol{#1}}

\begin{document}

\title{Line of Dirac monopoles embedded in a Bose-Einstein condensate}
\author{G.J.~Conduit}
\affiliation{Theory of Condensed Matter Group,
  Department of Physics, Cavendish Laboratory, J.J. Thomson Avenue,
  Cambridge, CB3 0HE, United Kingdom}

\date{\today}

\begin{abstract}
  The gauge field of a uniform line of magnetic monopoles is created
  using a single Laguerre-Gauss laser mode and a
  gradient in the physical magnetic field. We study the effect of these
  monopoles on a Bose condensed atomic gas, whose vortex structure
  transforms when more than six monopoles are trapped
  within the cloud. Finally, we study this transition with the collective
  modes.
\end{abstract}


\maketitle

A point source of magnetic flux, a magnetic monopole, has long been an
important missing component of grand unified and superstring
theories. Despite an exhaustive search for such
monopoles~\cite{Price75,Cabrera82}, no convincing evidence for the existence
of a stable magnetic monopole has been forthcoming. Analogies to certain
aspects of monopoles have been uncovered in physically tractable systems
including spin ice~\cite{Castelnovo08}, topological insulators~\cite{Qi09},
the anomalous quantum Hall effect~\cite{Fang03}, and superfluid
$^{3}\text{He}$~\cite{Blaha76}. Nevertheless, the experimental realization
of a true Dirac monopole remains an important target.

Over the past three years the cold atom gas has emerged as a promising new
system in which to explore the motion of particles in magnetic
fields~\cite{Dalibard11}. Starting with a seminal
experiment~\cite{Lin08,Lin09} that imposed an effective uniform magnetic
field on a Bose-Einstein condensate (BEC) of neutral atoms, the experimental
protocol has been extended to realize spin-orbit coupling~\cite{Lin11i}, and
effective electric fields~\cite{Lin11ii}. Following these successes, the
spotlight has recently turned to generating the high magnetic fields
required to study strongly correlated states such as the fractional quantum
Hall
effect~\cite{Jaksch03,Mueller04,Gerbier10,Sorensen05,Alba11,Cooper11,Aidelsburger11}.
However, cold atom gases also present a unique opportunity to study magnetic
field configurations not realizable in the solid state, in particular a
magnetic monopole~\cite{Moody86,Ruseckas05,Pietila09i,Pietila09ii}. The
experiments suggested to date demand either a laser setup of a superposition
of Laguerre-Gauss modes along with Hermite-Gauss
modes~\cite{Moody86,Ruseckas05,Pietila09i}, or three magnetic field
modes~\cite{Pietila09ii}, which have not yet been realized in
practice. Moreover, the configuration of a single monopole of fixed charge
means that there is no tuning parameter that would allow the direct
characterization of the state~\cite{Pietila09i}.

Here we propose a significantly simpler experimental protocol to embed a
uniform line of monopoles into a BEC. Motivated by previous
experiments~\cite{Lin09,Lin11i,Lin11ii} that successfully realized effective
gauge fields, our scheme adopts the same physical magnetic field gradient,
and replaces the co-propagating laser plane waves with the first optical
Laguerre-Gauss mode~\cite{Wright00,Olson07,John11}. The system undergoes a
transition with increasing monopole density, and we characterize the two phases
by their collective modes spectra.

\begin{figure}
 \begin{tabular}{p{0.5\linewidth}p{0.5\linewidth}}
  (a) Geometric setup&(b) Magnetic field realized\\[6pt]
  \includegraphics[width=0.8\linewidth]{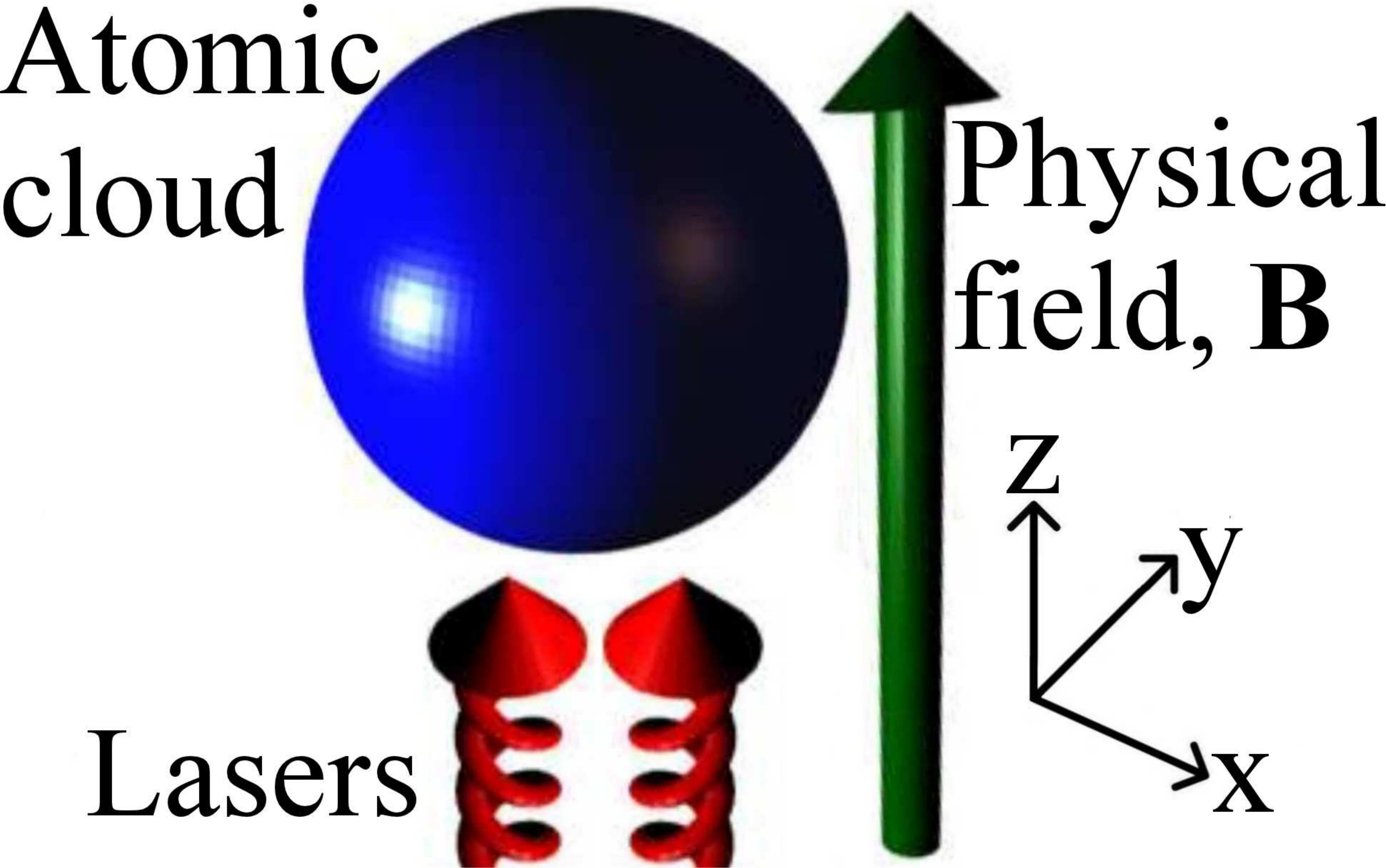}&\includegraphics[width=0.8\linewidth]{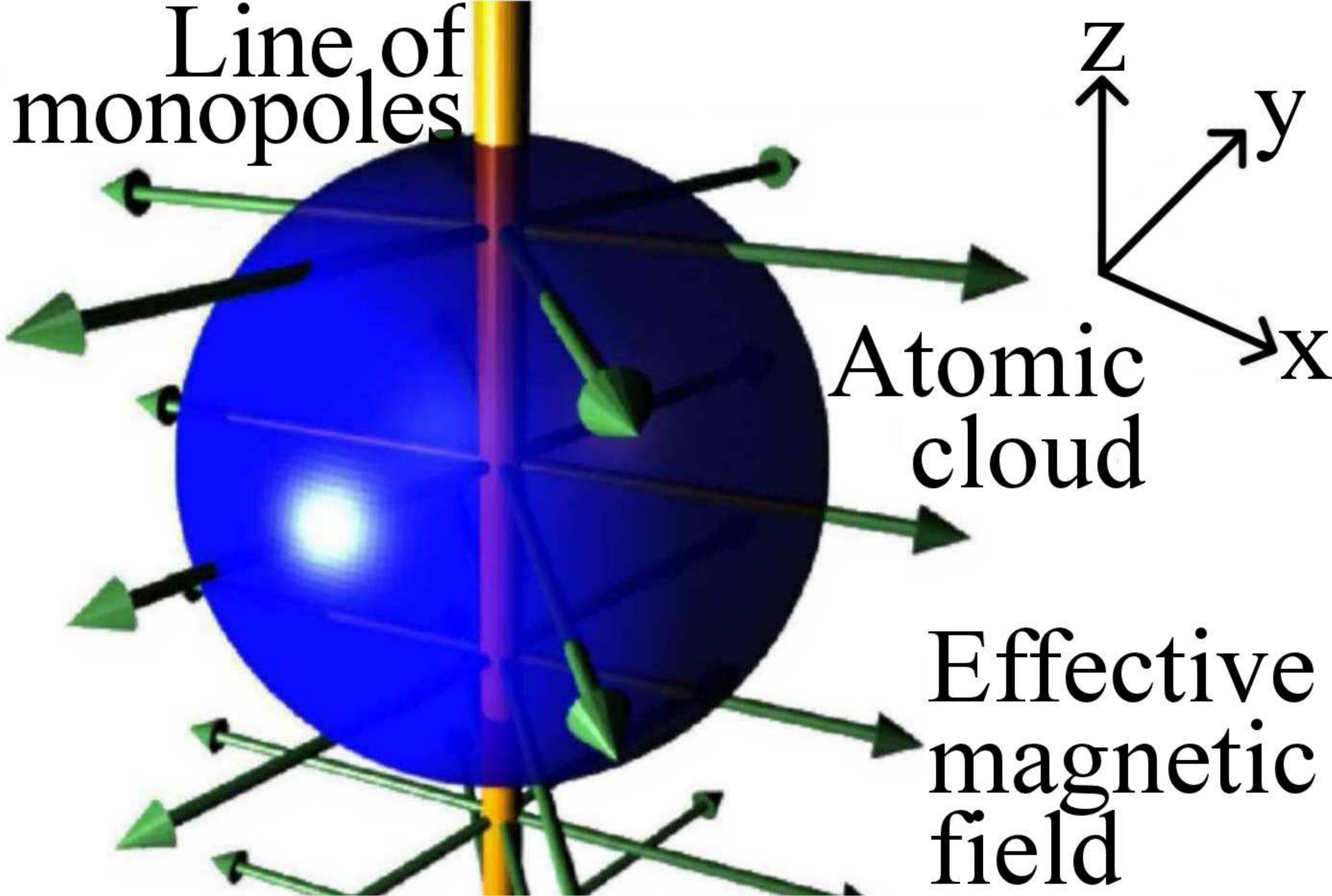}
 \end{tabular}
 \begin{tabular}{p{0.5\linewidth}p{0.5\linewidth}}
  (c) Atomic level scheme&(d) Magnetic field error\\[6pt]
  \includegraphics[width=0.9\linewidth]{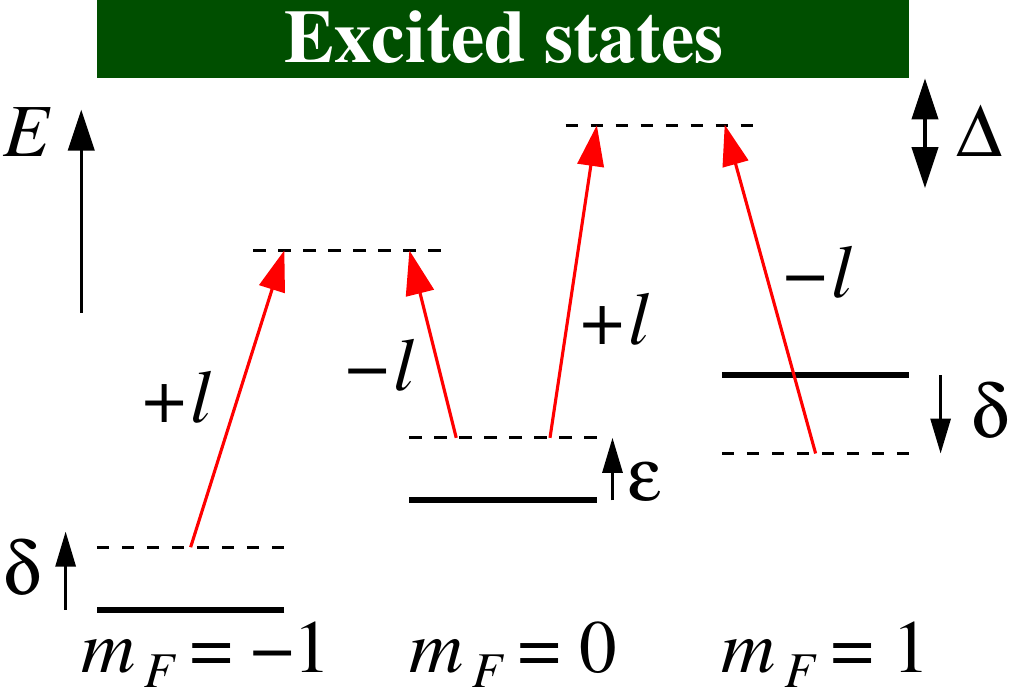}&\includegraphics[width=0.8\linewidth]{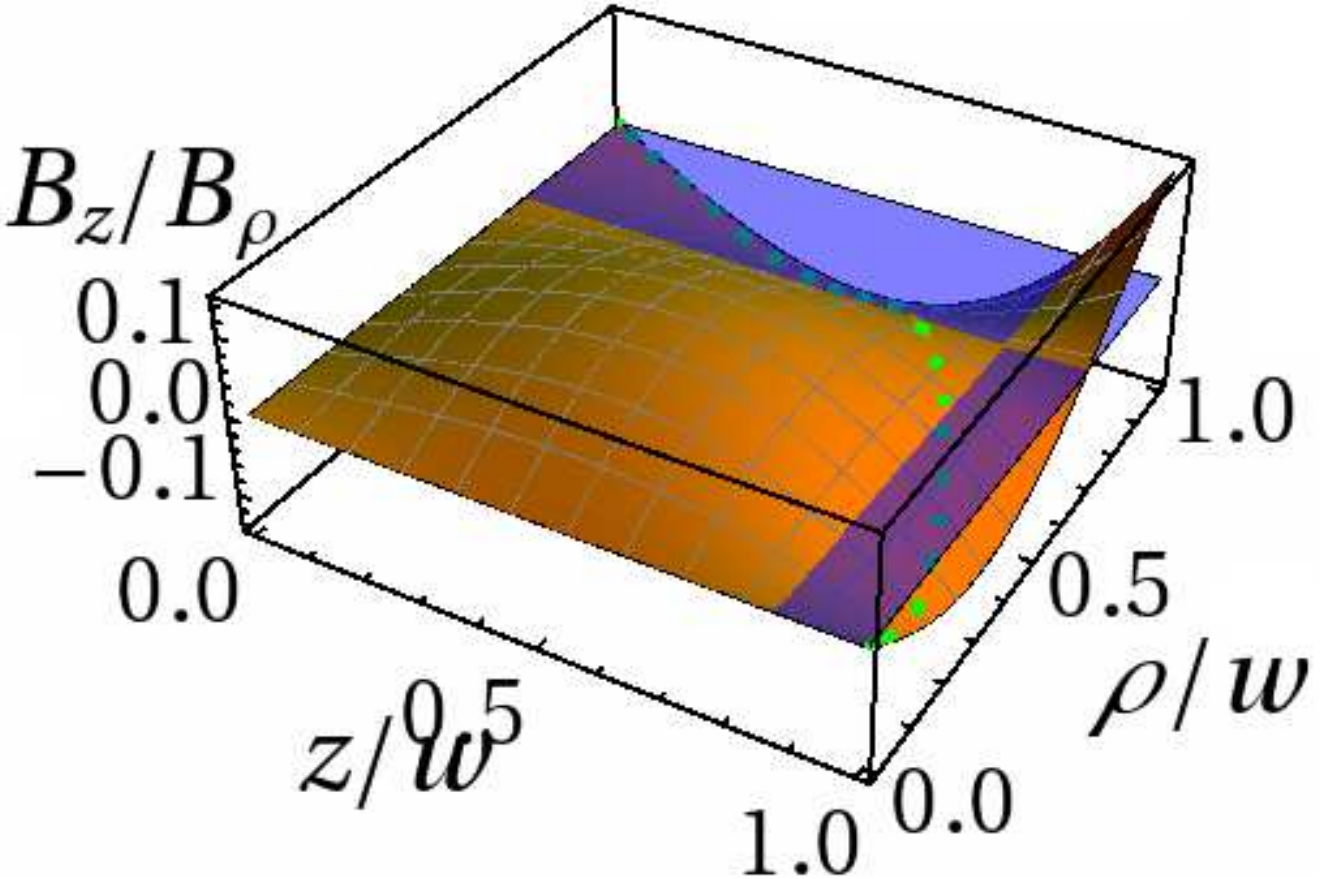}
 \end{tabular}
 \caption{(Color online) (a)~Geometry of the proposed scheme.
   Co-propagating laser beams with winding
   numbers $\pm\ell$ and the physical magnetic field
   $\vec{B}$ are aligned along $z$.  (b)~The gauge field realized, with the line of
   monopoles in orange, the effective magnetic field as green arrows, and the
   atomic cloud as the blue sphere. (c)~The level diagram for the three electronic
   ground states displaced by the external physical magnetic field $\vec{B}$.
   The irradiating laser beams $+\ell$ and $-\ell$ induce
   resonant coupling. In practice $\Delta\gg\delta$. (d)~The relative axial
   $B_{\text{z}}$ to radial magnetic field $B_{\rho}$. The dashed green line denotes
   the path of constant radius in the
   trap, and the blue sheet is $B_{\text{z}}/B_{\rho}=0$.}
 \label{fig:AtomicScheme}
\end{figure}

We first outline the experimental scheme to create the
gauge field of a line of monopoles. Starting from the geometry in
\figref{fig:AtomicScheme}(a), we trap bosonic atoms that have three coupled
internal levels (for example the $F=1$ manifold of $^{87}\text{Rb}$) labeled
by $m_{\text{F}}$. We apply a small physical magnetic field
$(B+B'z)\hat{\vec{z}}$ where the magnetic field gradient $B'$ could be
generated by an anti-Helmholtz coil pair. The field introduces a linear
Zeeman splitting so the atomic levels have
energies $g\mu_{\text{B}}m_{\text{F}}B+m_{\text{F}}\delta$,
where $\delta=g\mu_{\text{B}}B'z$ is the detuning from the Raman resonance
as shown in \figref{fig:AtomicScheme}(c). Co-propagating Laguerre-Gauss
laser beams induce the complex Rabi frequencies
$\kappa_{\pm}=\kappa_{0}(\sqrt{2}\rho/\sqrt{\ell}w)^{\ell}\exp(\ell/2-\rho^{2}/w^{2}\pm\cmplxi\ell\phi+\cmplxi
kz)$~\cite{Dalibard11}, with $\ell$ the beam winding number, $w$ the beam
width, $\rho=\sqrt{x^{2}+y^{2}}$ the distance from the $z$-axis, $\phi$ the
azimuthal angle, $\kappa_{0}$ the Rabi frequency at $\rho=w\sqrt{\ell/2}$,
and $k$ the plane wave vector. The single-photon detuning $\Delta$ from the
excited states manifold should be large versus the Rabi frequencies
$\Delta\gg\kappa_{0}$ so that the excited states can be adiabatically
eliminated and the Hamiltonian acts only on the ground state manifold.

The Hamiltonian in the basis set
$\{|m_{F}=-1\rangle,|m_{F}=0\rangle,|m_{F}=+1\rangle\}$ is
$\hat{H}=(-\hbar^{2}\nabla^{2}/2m+\hat{V}-\mu)\mat{1}+\hat{U}$, with $m$ the atomic mass, the
spin-independent trapping potential is
$\hat{V}\!=\!m\omega^2(\rho^{2}\!+\!z^{2})/2$ with $\omega$ the trap
curvature, and in the rotating wave approximation the coupling matrix is
\begin{align}
 \hat{U}=\frac{\hbar}{2}\left(
 \begin{array}{ccc}
  -2\delta&\kappa_{-}^{*}&0\\
  \kappa_{-}&\epsilon\kappa_{0}&\kappa_{+}\\
  0&\kappa_{+}^{*}&2\delta
 \end{array}
\right)\punc{,}
\end{align}
with $\epsilon$ the quadratic Zeeman shift in units of $\kappa_{0}$.  The
three bands of the Hamiltonian are split by $\sim\kappa_{0}\gg\delta$.
If the thermal energy and chemical potential $\mu$~\cite{Spielman09} are much lower
than $\hbar\kappa_{0}$~\cite{Lin09} then
only the ground state is occupied
\begin{align}
 |\chi\rangle\!=\!\e{2\cmplxi\ell\phi}\cos^{2}\frac{\theta}{2}|-1\rangle
             \!-\!\frac{\sin\theta}{\sqrt{2}}|0\rangle
             \!+\!\e{-2\cmplxi\ell\phi}\sin^{2}\frac{\theta}{2}|1\rangle\punc{,}
\end{align}
with mixing angle $\tan{\theta}=|\kappa_{+}|/\sqrt{2}\delta$, and energy
$-\hbar\sqrt{\delta^{2}+\kappa_{0}^{2}/2}$.

\begin{figure}
 \begin{tabular}{ccccc}
  &\multicolumn{2}{l}{(a) $N=4$, realized}&\multicolumn{2}{l}{(b) $N=12$, realized}\\[-2pt]
   \multirow{2}*[0.02\linewidth]{\includegraphics[width=0.14\linewidth]{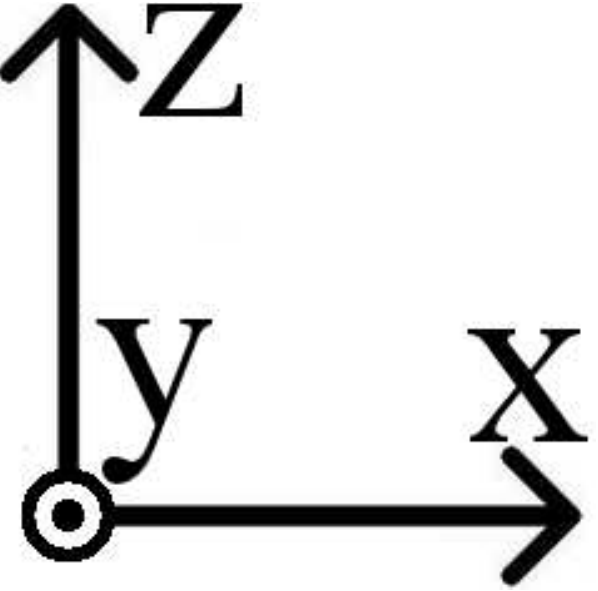}}&
   \multirow{2}*[0.08\linewidth]{\includegraphics[width=0.24\linewidth]{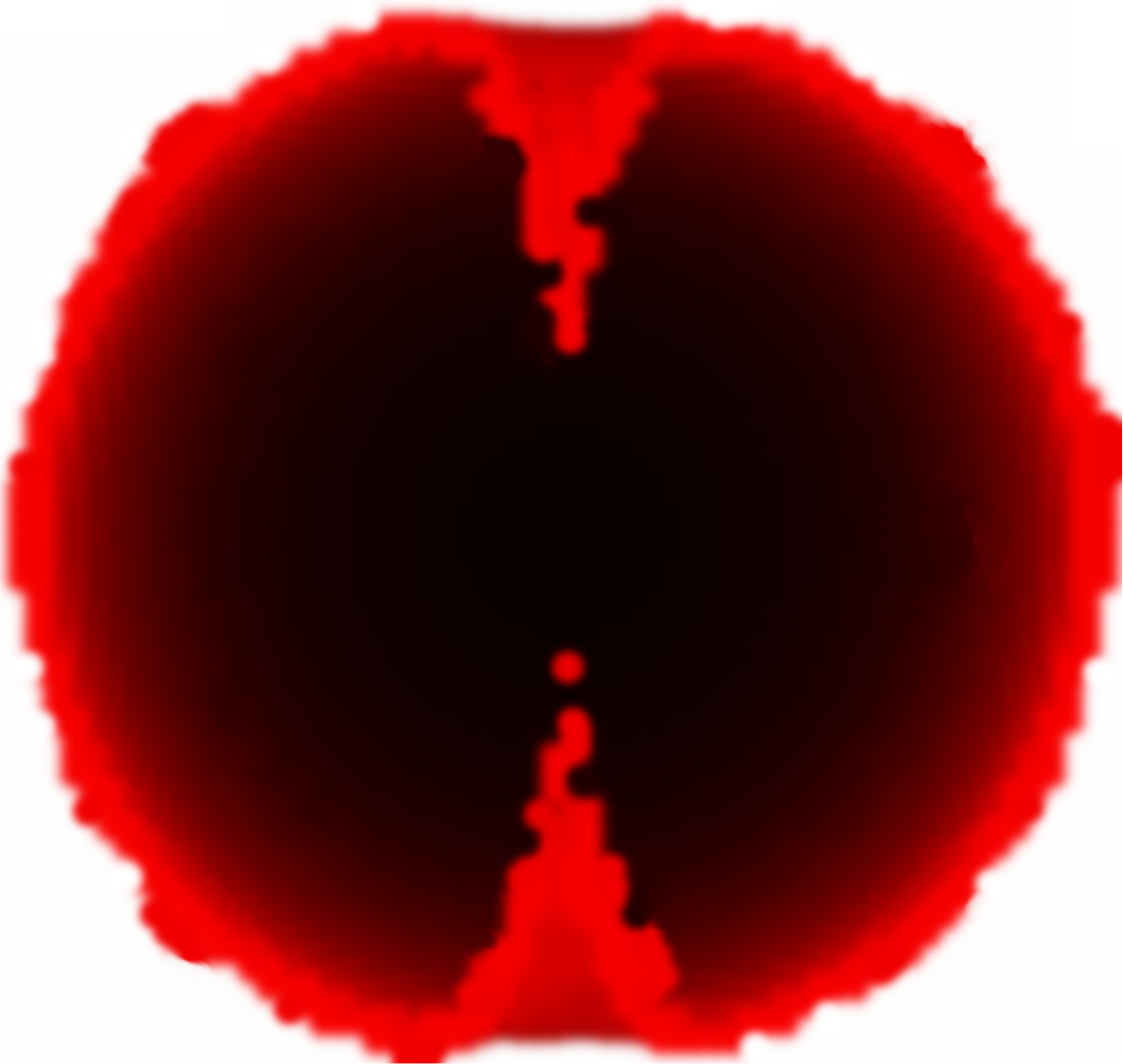}}&
   \includegraphics[width=0.15\linewidth]{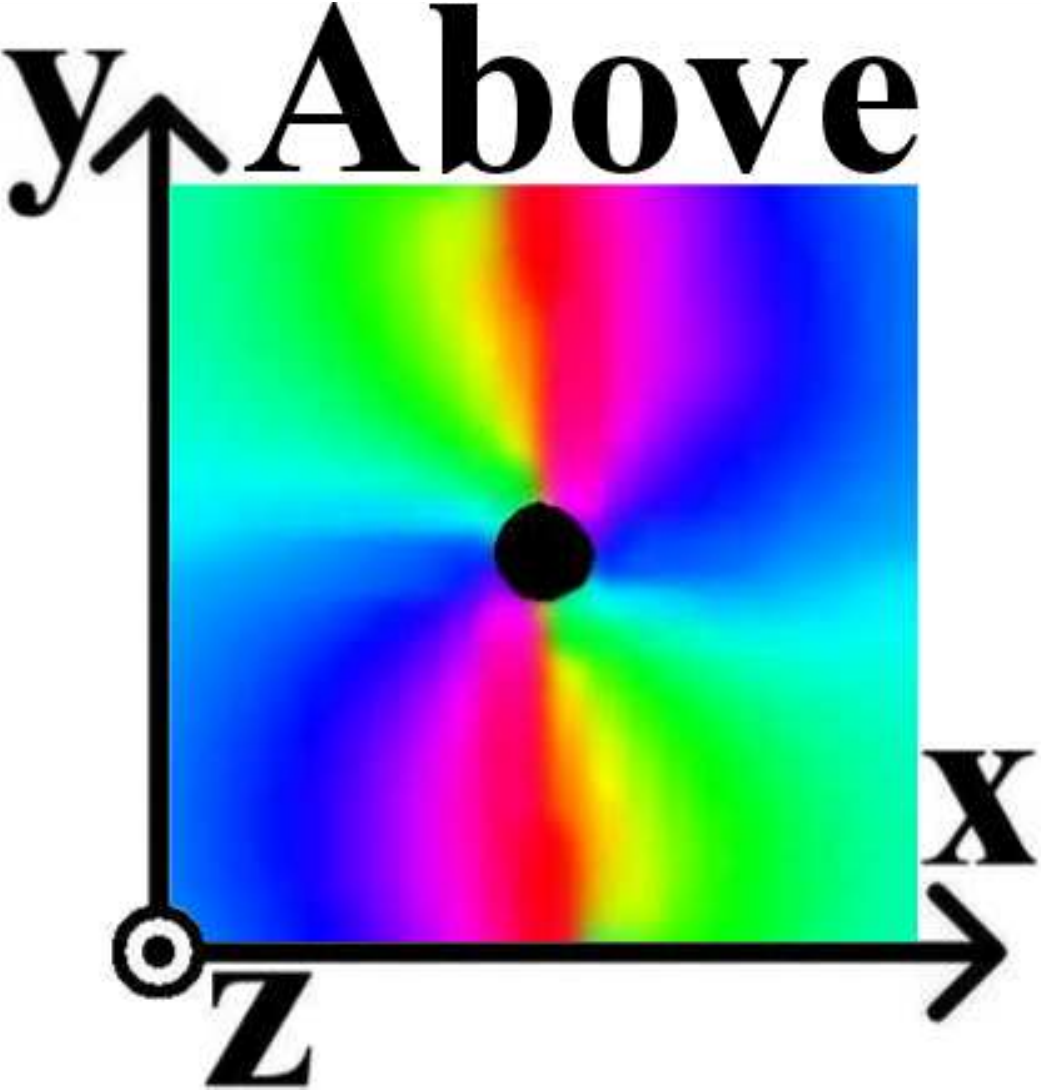}&
   \multirow{2}*[0.08\linewidth]{\includegraphics[width=0.24\linewidth]{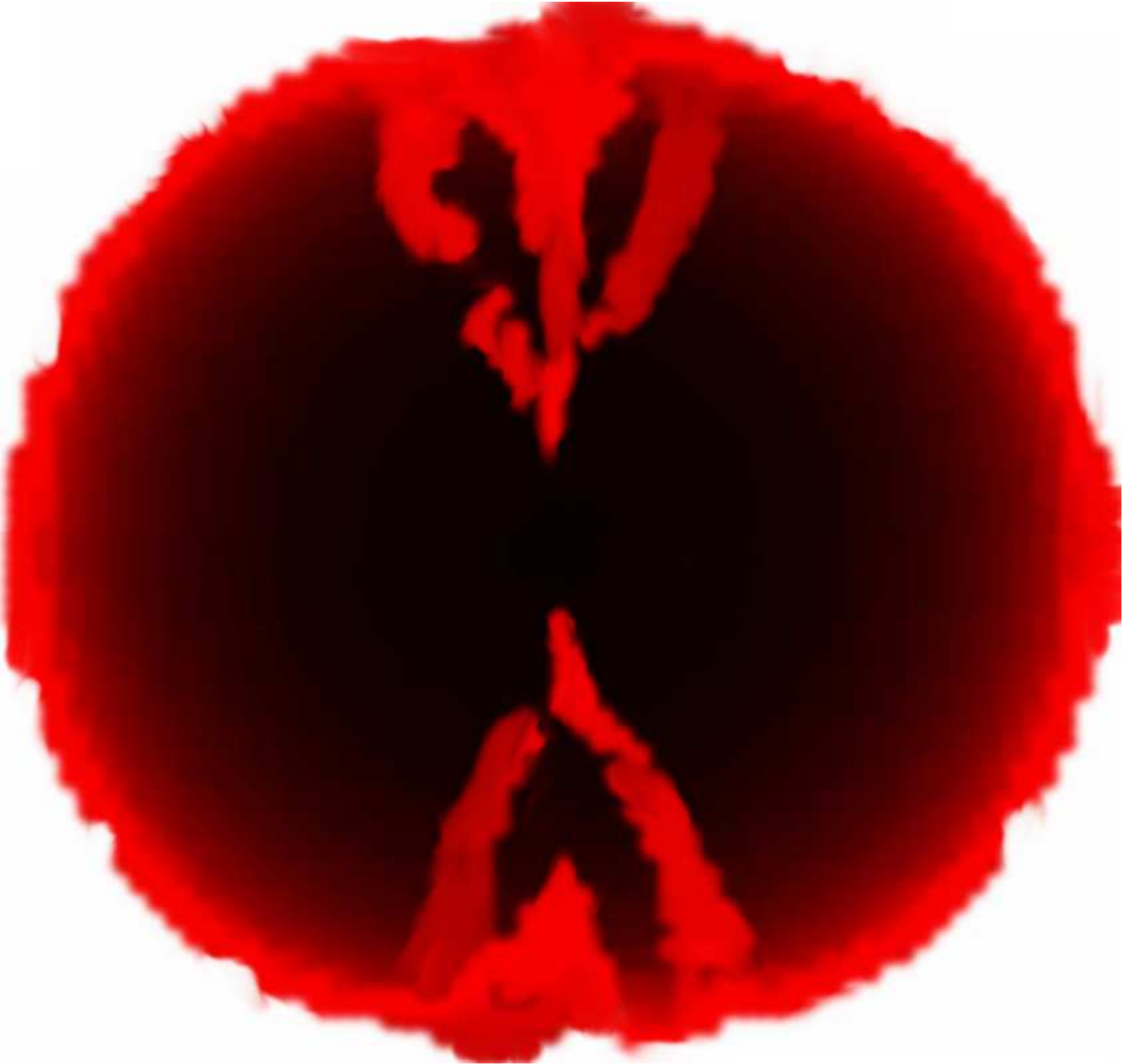}}&
  \includegraphics[width=0.15\linewidth]{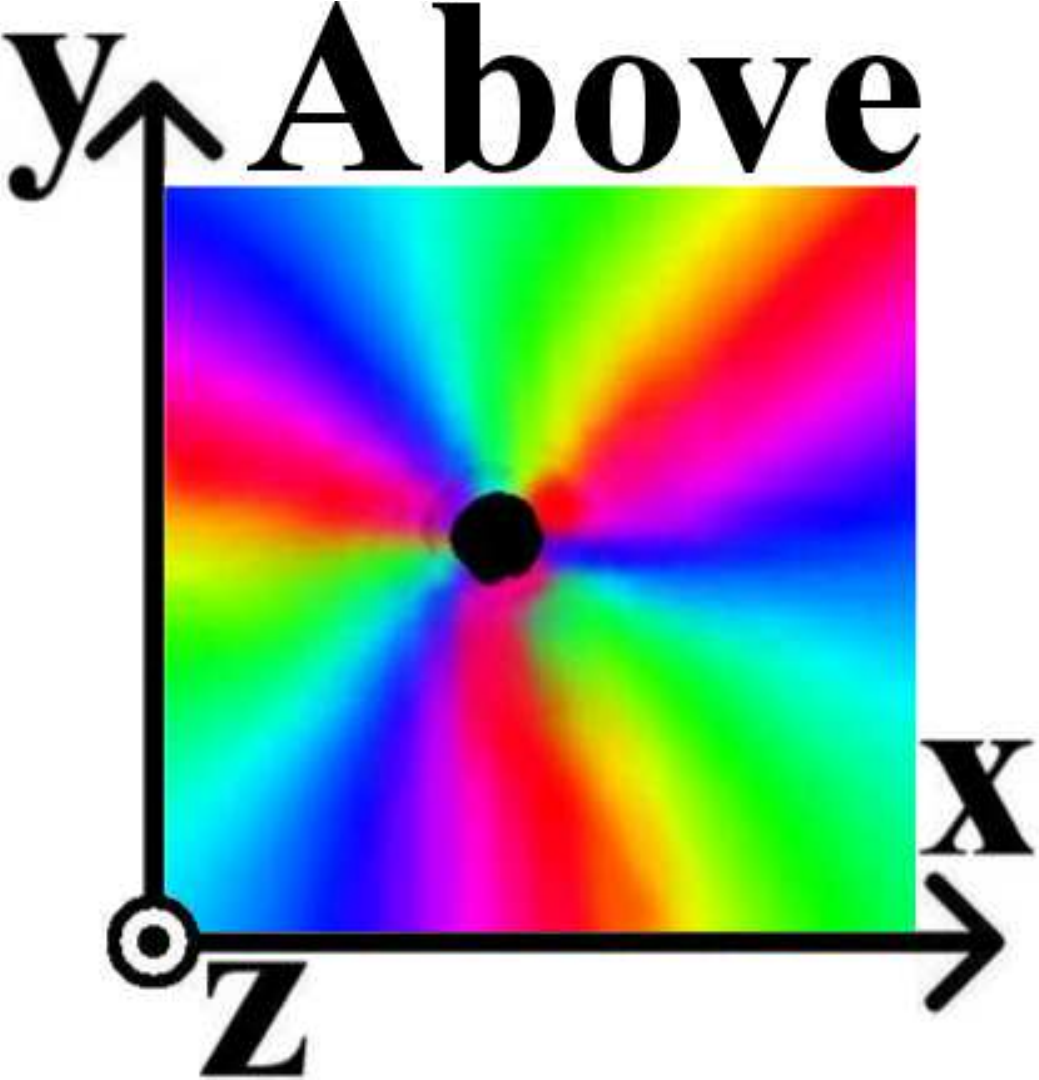}\\[-3pt]
   &&\includegraphics[width=0.15\linewidth]{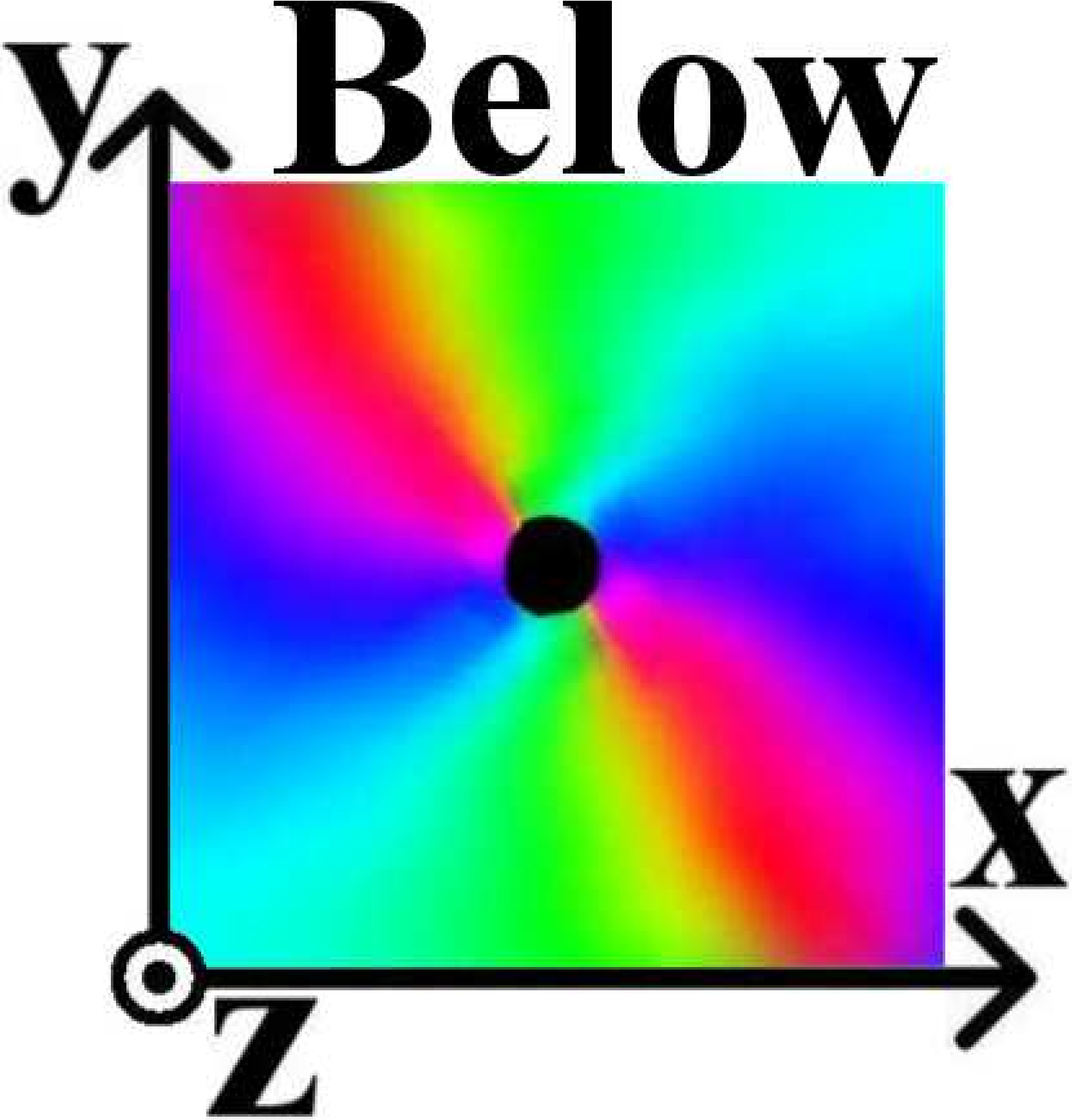}&&
   \includegraphics[width=0.15\linewidth]{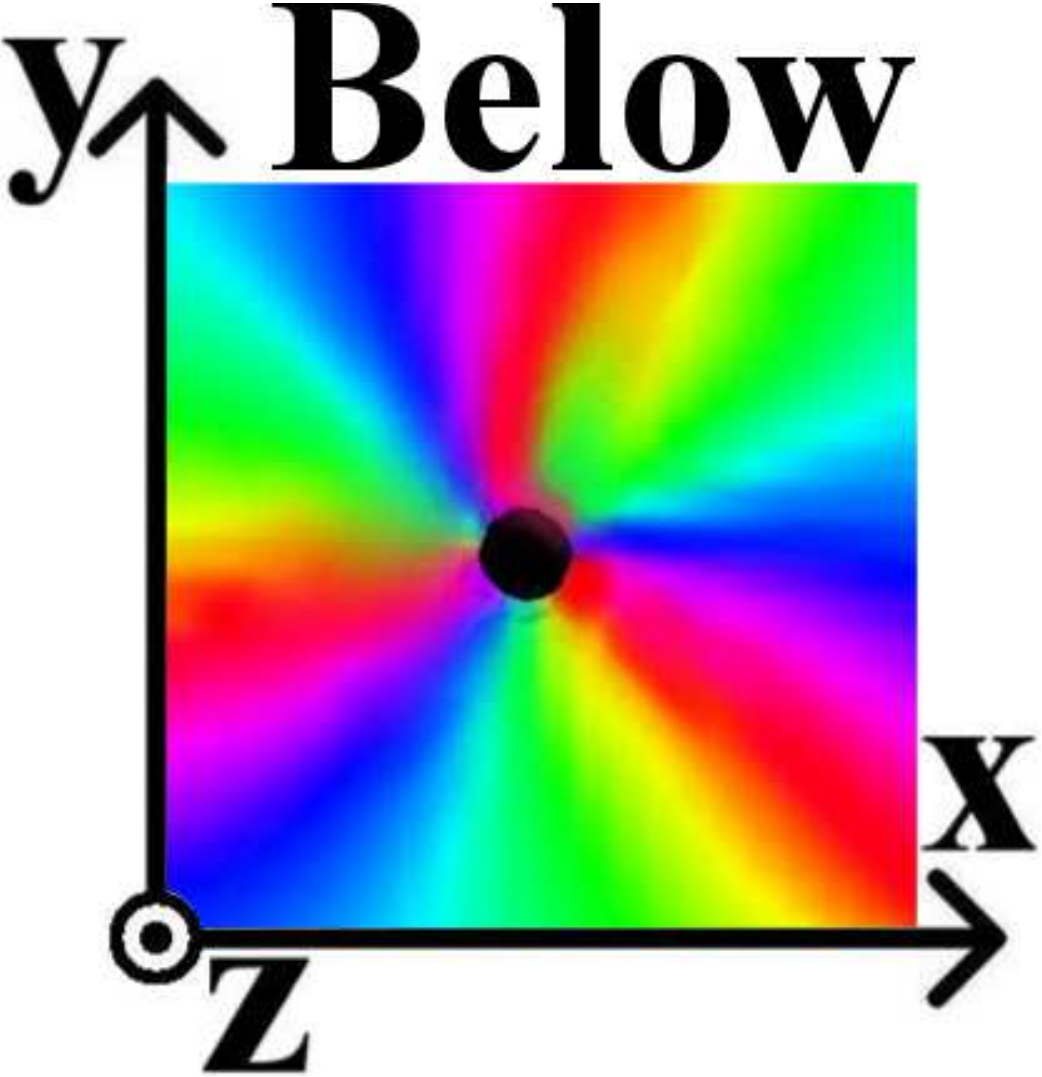}
  \end{tabular}
  \begin{tabular}{cccc}
  (c) $N=55$&(d) $N=4$&(e) $N=12$&(f) $N=55$\\[-2pt]
  realized&idealized&idealized&idealized\\[-2pt]
  \includegraphics[width=0.24\linewidth]{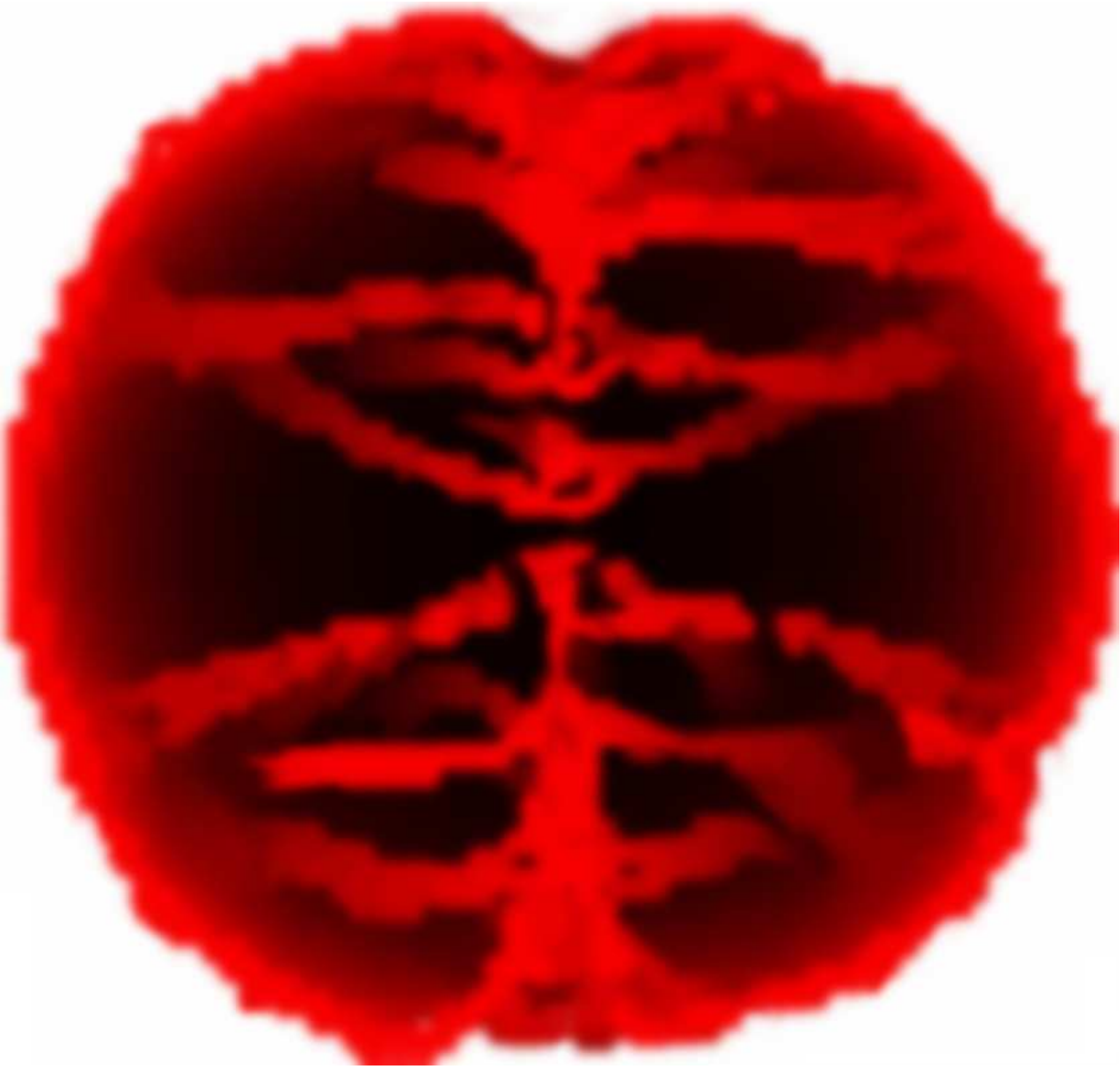}&
  \includegraphics[width=0.24\linewidth]{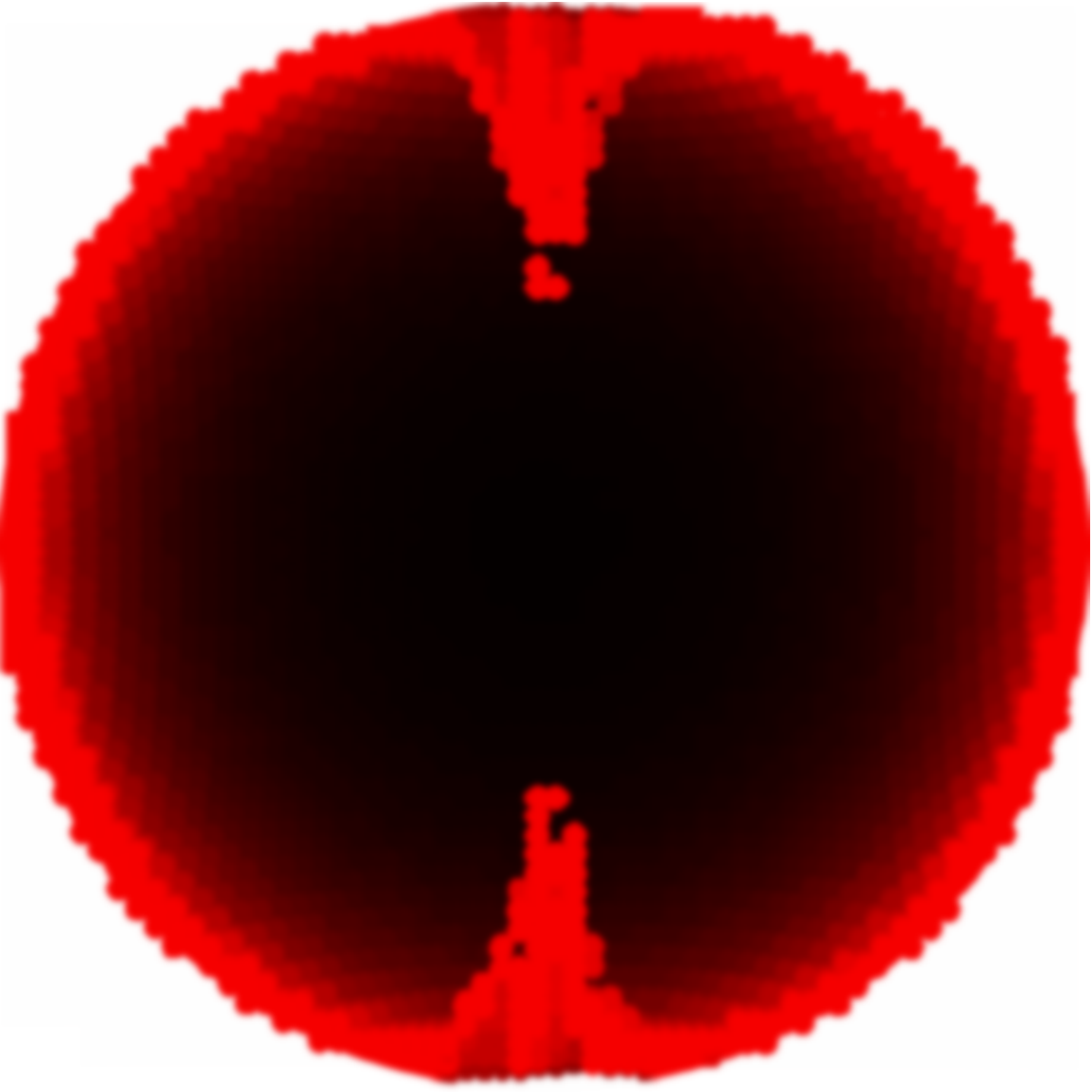}&
  \includegraphics[width=0.24\linewidth]{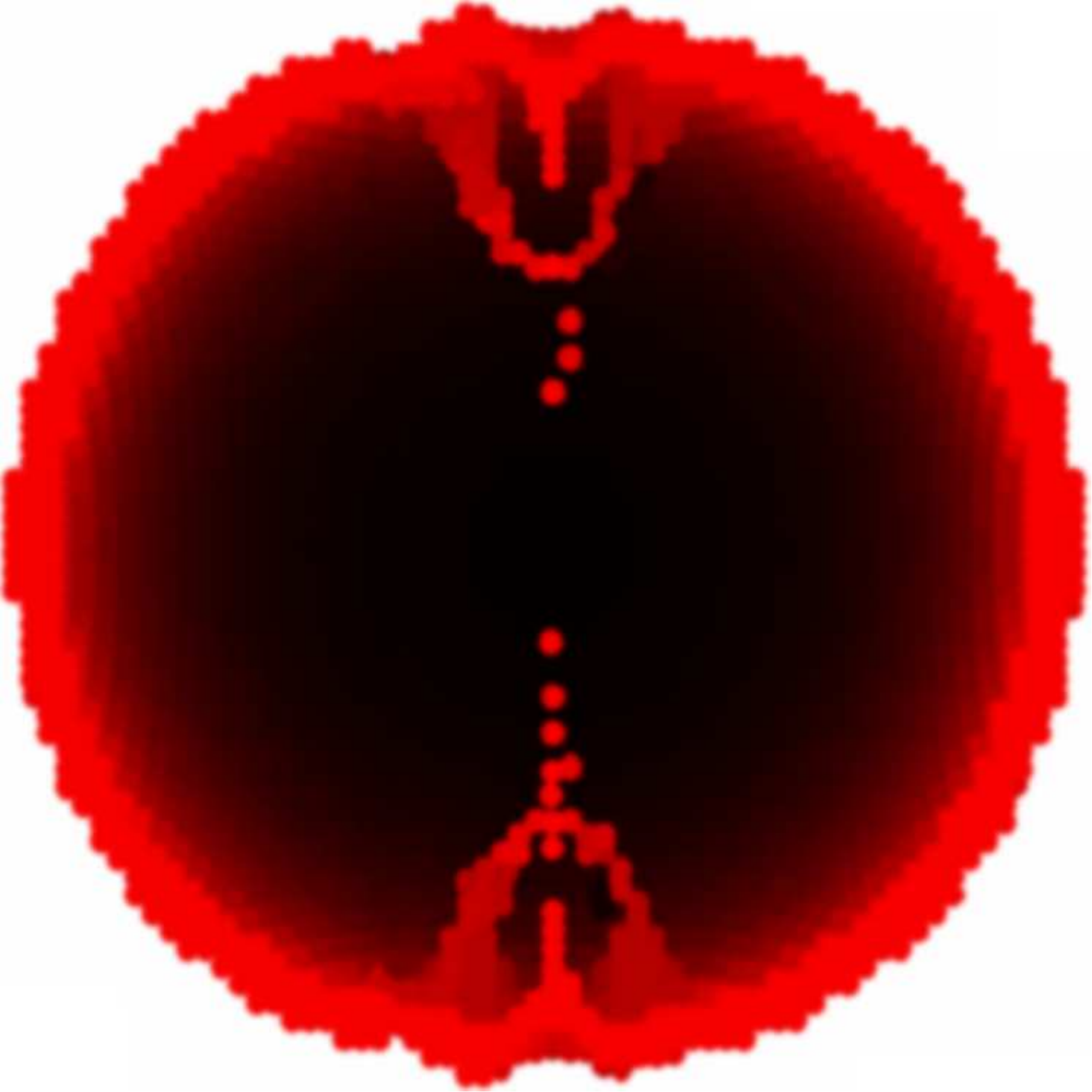}&
  \includegraphics[width=0.24\linewidth]{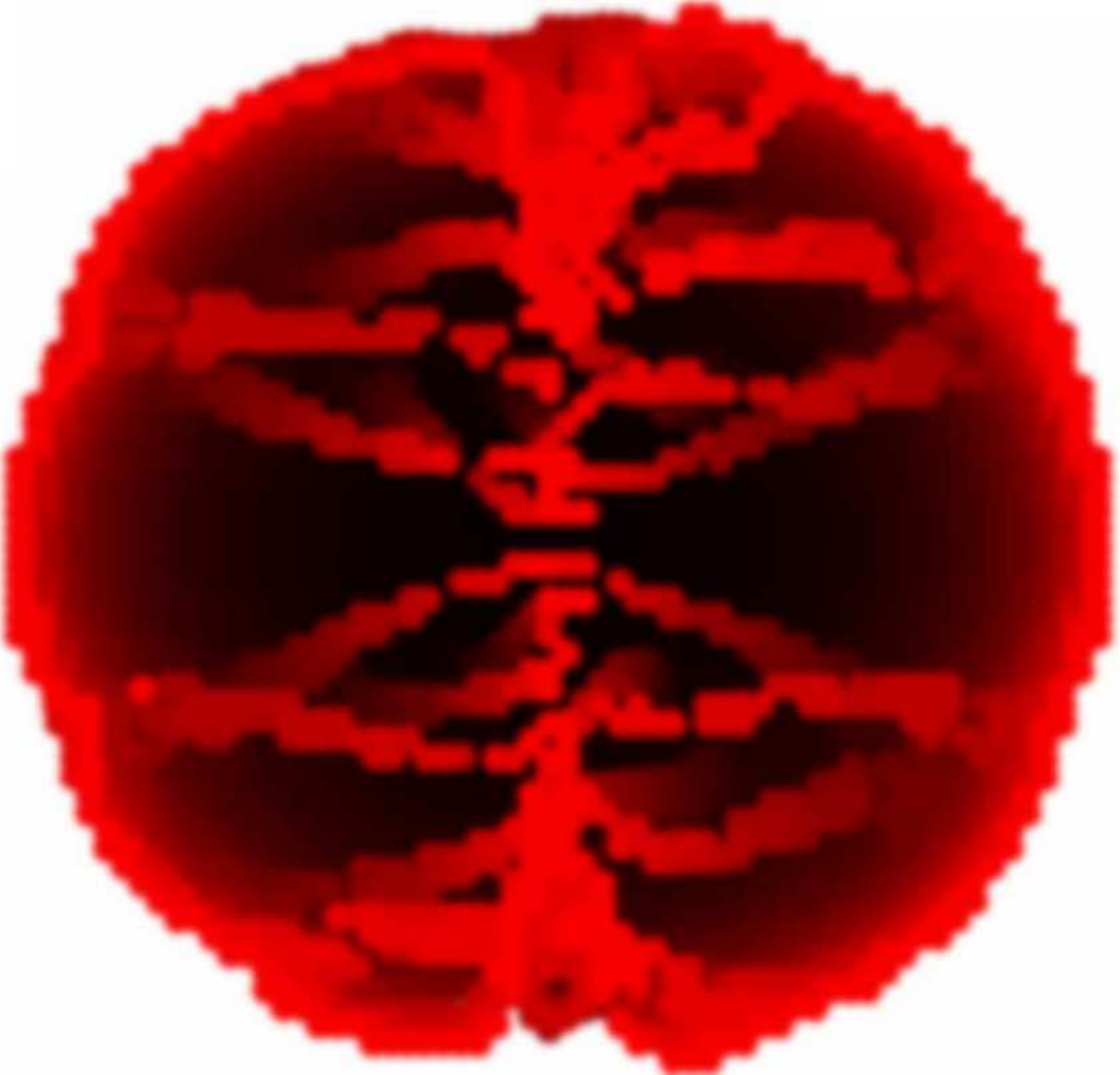}
 \end{tabular}
 \caption{(Color online) The particle density isosurfaces of BECs
   containing (a) $N=4$, (b) $N=12$, (c) $N=55$ monopoles for the gauge field
   of the proposed setup. The isosurfaces for the artificial situation of a
   perfect line of (d) $N=4$, (e) $N=12$, and (f)
   $N=55$ monopoles. The brightness denotes
   the depth of the site on the $y$-axis. The
   right-hand plots of (a,b) view the condensate along the
   $z$-axis, and use the hue to display the condensate phase.}
 \label{fig:DensityPlots}
\end{figure}

Following the prescription of Ref.~\cite{Dalibard11}, we can calculate the
gauge field as
$\vec{A}=2\cmplxi\hbar\langle\chi|\nabla\chi\rangle=\hbar\ell\cos\theta\hat{\vecgrk{\phi}}/\rho$.
The corresponding effective magnetic field is $\vec{B}=\nabla\times\vec{A}$
so
\begin{align}
 \!\!\!\left(\!\!\!\begin{array}{c}
 B_{\rho}\\B_{\phi}\\B_{\text{z}}
 \end{array}\!\!\!\right)
 \!\!=\!\frac{\hbar\ell\delta'}
 {\kappa\rho[1\!+\!(\!\frac{z\delta'}{\kappa_{0}}\!)^{\!2}]^{3/2}}
 \!\!\left(
 \begin{array}{c}
  1\\0\\\!\!\!\frac{z}{\sqrt{\ell}w}\!\!\left[\!\alpha\!+\!\beta\frac{z^2}{w^2}\!\right]
  \!\!\left[2\!\left(\!\frac{\rho}{w}\!\right)^{2}\!\!-\!\ell\right]\!\!
 \end{array}
 \right)\!\!\punc{,}
 \label{eqn:TrueB}
\end{align}
where $\alpha\!=\!\sqrt{2}\!+\!\epsilon\!+\!3\epsilon^2/10$ and 
$\beta\!=\!1/\sqrt{2}\!+\!\epsilon/2\!+\!17\epsilon^2/128$.
The scalar potential defined by
$W\!=\!\hbar^{2}[(\vecgrk{\nabla}\theta)^2\!+\!\sin^{2}\theta(\vecgrk{\nabla}\phi)^2]/8m$ is
\begin{align}
 W&=\frac{\hbar^{2}}{2m}\frac{1}{\sqrt{1+\left(\frac{\delta'z}{\kappa_{0}}\right)^{2}}}
 \Biggl\{
 \frac{1}{\sqrt{1+\left(\frac{\delta'z}{\kappa_{0}}\right)^{2}}}\left(\frac{\ell}{\rho}\right)^{2}\neweqnline
 &+\left(\frac{\delta'}{2\kappa_{0}}\right)^{2}
 \left[1+\left(\frac{z}{\rho}\right)^{2}\left(1-2\left(\frac{\rho}{w}\right)^{2}\right)^{2}\right]\Biggr\}
 \punc{.}
\end{align}
After expressing the magnetic field gradient through $\delta=\delta'z$, we
expand these expressions in the $\kappa_{0}\gg\delta'w$ limit, and about the
radius $\rho_{\text{max}}=w\sqrt{\ell/2}$ of maximum laser beam intensity to
yield the effective magnetic field
\begin{align}
 \left(\begin{array}{c}
 B_{\rho}\\B_{\phi}\\B_{\text{z}}
 \end{array}\right)\approx\frac{\sqrt{2}\hbar\ell\e{\ell}\delta'}{\kappa_{0}\rho}
 \left(
 \begin{array}{c}
  1\\0\\0
 \end{array}
 \right)\punc{,}
\end{align}
and in the same limit the scalar potential is
$W=\hbar^{2}(\ell/\rho)^{2}/2m$. The effective magnetic field formed has the
configuration shown in \figref{fig:AtomicScheme}(b): it is directed radially
and decays as $1/\rho$, so it represents the magnetic field emanating from a
line of Dirac monopoles along $\hat{\vec{z}}$, with line density
$2\sqrt{2}\ell\e{\ell}\delta'/\kappa_{0}$. However, from \eqnref{eqn:TrueB},
$B_{\text{z}}=0$ only exactly at the radius
$\rho_{\text{max}}=w\sqrt{\ell/2}$, but remains small away from it.
To minimize $B_{\text{z}}$ over all radii we set $\epsilon=-2.8$, which
yields the variation of $B_{\text{z}}/B_{\rho}$ shown in
\figref{fig:AtomicScheme}(d).  Over the range of $\rho$ and $z$ within a
trap (in \figref{fig:AtomicScheme}(d) set so $\rho^2+z^2<w^2$) we find that
$|B_{\text{z}}/B_{\rho}|<0.1$ and so the effective field closely
approximates that of a line of monopoles. 
We also note that the anti-Helmholtz coil pair necessarily generates a radial
physical magnetic field gradient as well as the $B'$ along the $z$-axis. This
generates a $B_{\text{z}}$ component that could
also be removed by the quadratic Zeeman shift. In both cases as $\vec{A}$ is
continuous away
from $\rho=0$ flux is conserved and the only monopoles lie along the
$z$-axis. The $B_{\text{z}}$ component of the effective magnetic field is
then a small field component
corresponding to a changing line density of monopoles. This will be further
tested in \figref{fig:DensityPlots} by comparing the BEC ground state for
both gauge fields.

To probe the experimental prospects of the proposed configuration we first
check the maximum achievable monopole density.  For $\ell=1$, using the
realistic~\cite{Lin09} values $\delta'=4.5E_{\text{r}}/\mu\text{m}$ and
$\kappa_{0}=16E_{\text{r}}$, with $E_{\text{r}}$ the recoil energy, a cloud
of diameter $d=80\mu\text{m}$ contains
$N=2\sqrt{2}\e{}\delta'd/\kappa_{0}\approx80$ monopoles. We adopt the
natural definition of monopole charge density so that the flux quantum $\Phi_{0}$
emanating from a single charge winds the phase of the BEC by a factor of
$2\pi$.

Having formed the effective gauge field of a line of monopoles,
we now develop the formalism to study its influence on a BEC. To seek and
study the ground state we discretize~\cite{Pietila09i,Pietila09ii} the
Hamiltonian onto a lattice of spacing $a$. The derivative at a lattice site
$\vec{R}=\{x,y,z\}$ along a single link $\vec{r}$ is transformed to
$[D_{\vec{r}}\psi(\vec{R})]^{\dagger}[D_{\vec{r}}\psi(\vec{R})]\to
(\psi_{\vec{R}+\vec{r}}-U_{\vec{R},\vec{r}}\psi_{\vec{R}})^{\dagger}
(\psi_{\vec{R}+\vec{r}}-U_{\vec{R},\vec{r}}\psi_{\vec{R}})/a^{2} \equiv
C_{\vec{R},\vec{r}}$, where the hopping matrix is
$U_{\vec{R},\vec{r}}=\exp(\cmplxi\vec{A}(\vec{R})\cdot\vec{r}/\hbar)$, and
$\psi_{\vec{R}}$ is the condensate order parameter at $\vec{R}$. With this
in place the mean-field energy is
\begin{align*}
 E\!=\!\frac{\hbar^{2}}{2m}\!\sum_{\vec{R},\vec{r}}\!C_{\vec{R},\vec{r}}
 \!+\!\sum_{\vec{R}}\!
\left[(V_{\vec{R}}\!+\!W_{\vec{R}}\!-\!\mu)|\psi_{\vec{R}}|^{2}
\!+\!\frac{c}{2}|\psi_{\vec{R}}|^{4}\right]\!\punc{,}
\end{align*}
with external potential
$V_{\vec{R}}\!=\!m\omega^{2}(x^{2}\!+\!y^{2}\!+\!z^{2})/2$, and a chemical
potential $\mu$ to fix the number of particles. We minimize the free energy
variationally with respect to $\psi_{\vec{R}}^{*}$ to yield the discrete
Gross-Pitaevskii (GP) equation
\begin{align}
 \mu\psi_{\vec{R}}=&-\frac{\hbar^{2}}{2ma^{2}}\sum_{\vec{r}}
 \left(U_{\vec{R},\vec{r}}^{*}\psi_{\vec{R}+\vec{r}}-2\psi_{\vec{R}}
 +U_{\vec{R}-\vec{r},\vec{r}}\psi_{\vec{R}-\vec{r}}\right)\neweqnline
 &+\left(V_{\vec{R}}+W_{\vec{R}}+c|\psi_{\vec{R}}|^{2}\right)\psi_{\vec{R}}\punc{.}
\end{align}
To calculate the ground state we solve the GP equations iteratively using
the over-relaxation scheme on a grid of $200\times200\times200$ sites with
periodic boundary conditions.

With a tool to determine the ground state wave function in place, to calculate
the collective mode spectrum we adopt the linear-response
theory~\cite{Edwards96} that was successfully used to study the excitation
frequencies of a uniform BEC in a vortex state~\cite{Dodd97}. We first
substitute the general form for the solution of the Gross-Pitaevskii
equation
\begin{align}
 \Psi_{\vec{R},t}=\e{\cmplxi\mu t/\hbar}
 \left[\psi_{\vec{R}}+u_{\lambda,\vec{R}}\e{-\cmplxi\Omega_{\lambda}t}
 +v_{\lambda,\vec{R}}^{*}\e{\cmplxi\Omega_{\lambda}t}\right]\punc{,}
\end{align}
into the Hamiltonian. Here $\psi_{\vec{R}}$ is the ground state wave
function and $\lambda$ is the index of the eigenvalue
$\Omega_{\lambda}$ with eigenfunctions $u_{\lambda,\vec{R}}$ and
$v_{\lambda,\vec{R}}$. This recovers the time-independent Gross-Pitaevskii
equation for $\psi_{\vec{R}}$ along with the linear-response Bogoliubov
equations
\begin{align}
 \left(\begin{array}{cc}
 \mathcal{L}_{\vec{R}}&-\frac{c}{2}|\psi_{\vec{R}}|^{2}\\
 \frac{c}{2}|\psi_{\vec{R}}|^{2}&-\mathcal{L}_{\vec{R}}
 \end{array}\right)
 \left(\begin{array}{c}
 u_{\lambda,\vec{R}}\\v_{\lambda,\vec{R}}
 \end{array}\right)=\hbar\Omega_{\lambda}
 \left(\begin{array}{c}
 u_{\lambda,\vec{R}}\\v_{\lambda,\vec{R}}
 \end{array}\right)\punc{,}
\end{align}
with $\mathcal{L}_{\vec{R}}=-\hbar^{2}
\sum_{\vec{r}}(U^{*}_{\vec{R}-\vec{r},\vec{r}}-2+U_{\vec{R},\vec{r}})/2ma^2
+V_{\vec{R}}+W_{\vec{R}}-\mu+c|\psi_{\vec{R}}|^{2}$.  Using the condensate
order parameter for the ground state, we use the Arnoldi method to solve
the Bogoliubov equations to find the lowest eigenstates and eigenmodes that
are the collective modes.

\begin{figure}
 \includegraphics[width=0.9\linewidth]{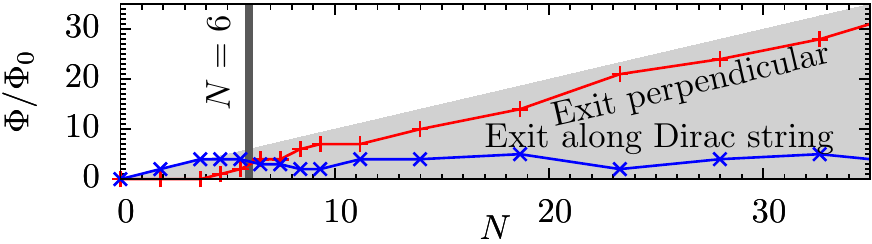}
 \caption{(Color online) Flux $\Phi$ exiting the cloud as a function of the
   number of embedded monopoles    $N$. The red line $(+)$ shows the flux exiting
   radially, the blue line $(\times)$ the flux
   exiting along the Dirac string. The grey shading denotes the region where
   the escaping flux is necessarily less than the monopolar charge
   contained. The vertical black line shows the theoretically predicted
   crossover when flux first exits away from the string.}
 \label{fig:PhaseDiag}
\end{figure}

With the physical setup and numerical
procedure in place, we now study the ground state configuration of our
trapped BEC embedded with a line of monopoles. We first examine the phase
behavior of the ground state with increasing monopole density. Second, to
fully characterize the state, we calculate their collective mode spectrum.
We study $100\mathord{,}000$ atoms in a spherical trap
with $\omega=0.1$ and interaction strength $c=0.05$ in atomic units
$\hbar=m=1$. The Laguerre-Gauss beams have winding
number $\ell=1$ and beam width $w=40a$, and vary the detuning gradient
$\delta'/\kappa_{0}$ to control the monopolar charge density,
which within a characteristic length $w$ of the center is
$4\sqrt{2}\e{}\delta'w/\kappa_{0}$. We use the realistic
gauge field $\vec{A}=2\hbar\cos\theta\hat{\vecgrk{\phi}}/\rho$.

With no embedded monopoles the BEC is spherically symmetric in the trap.  In
\figref{fig:DensityPlots}(a) we examine the density isosurfaces for a cloud
that encloses $N=4$ monopolar charges. The density falls along the $z$-axis
and the phase of the condensate order parameter winds by $4\pi$ around the
$z$-axis showing that the flux escapes symmetrically through the top and
bottom of the cloud forming vortices. On increasing the number of trapped
monopoles to $N=12$, in \figref{fig:DensityPlots}(b) now three flux quanta
pass through the top of the cloud, and a further three through induced
vortices directed away from the axis in the upper hemisphere, with the
situation repeated in the lower hemisphere. The stepwise change where flux
starts to escape away from the $z$-axis occurs at $N\approx6$. The
cylindrical symmetry of the cloud is broken by the random site updates on
the computational lattice.  When a large number of monopoles $N=55$ are
trapped by the cloud in \figref{fig:DensityPlots}(c) we observe that the
vast majority of the flux escapes through the side of the cloud broadly
symmetrically. Finally, we compare the predictions for the gauge field
realized by the proposed laser setup,
$\vec{A}=2\hbar\cos\theta\hat{\vecgrk{\phi}}/\rho$
(\figref{fig:DensityPlots}(a,b,c)), with an idealized
$\vec{A}=4\e{}\hbar\delta\hat{\vecgrk{\phi}}/\kappa_{0}\rho$ of the
artificial situation of the magnetic field of a perfect line of monopoles
(\figref{fig:DensityPlots}(d,e,f)). The distribution of the vortices formed
by the exiting flux are broadly similar. Therefore the gauge field realized
by the laser setup accurately models that of an ideal line of monopoles. We
also checked the stability of the results while changing the random number
seed of the lattice updates, boundary conditions between periodic and hard
wall, the orientation of the line of monopoles relative to the lattice, and
lattice size.

Having studied clouds containing
different monopolar charges, we now summarize their behavior in
\figref{fig:PhaseDiag}. In particular we focus on the route of the escaping
flux, which is along the $z$-axis for $N\lesssim5$ and penetrates through
the side walls of the cloud for $N\gtrsim8$. When $N\gtrsim8$, the flux still
exits along the $z$-axis, but remains fewer than a total of $\sim6$ flux
quanta. To understand why there is a transition in behavior at $5\lesssim
N\lesssim8$ we can compare the energy cost of flux exiting the cloud through
the two different routes. The energy of a vortex per unit length is
$C\gamma^{2}$~\cite{Bruun01} where $\gamma$ is the number of circulation
quanta of the vortex and $C$ a constant defined in~\cite{Bruun01}.  For a
system of radius $R$, length $L$, and monopole line density $\eta$, the
energy of all the flux escaping along the $z$-axis is $C\eta^{2}L^{3}/12$,
whereas if the flux exits radially the total energy is $C\eta LR$. Therefore
the flux exits radially when $12R>\eta L^{2}$. Here the cost of
multiple flux quanta exiting along the same vortex line out of the
top or bottom of the cloud is outweighed by the energy cost of flux exiting
through longer vortex lines out of the cloud sides.  For a spherically
symmetric trap where $L=2R$ and $N=\eta L$, the flux exits radially when
$N\ge6$. This can be readily seen in \figref{fig:PhaseDiag} where flux exiting
radially rises rapidly when more than six flux quanta are enclosed by the
cloud. Here, as in previous studies of rotation-induced vortices in BECs,
the vortices will be too small to image directly. However, if the cloud is
ballistically expanded the vortices are magnified and could be
imaged~\cite{Dalibard00,Shaeer01,Engels02,Coddington03}.

\begin{figure}
 \includegraphics[width=0.95\linewidth]{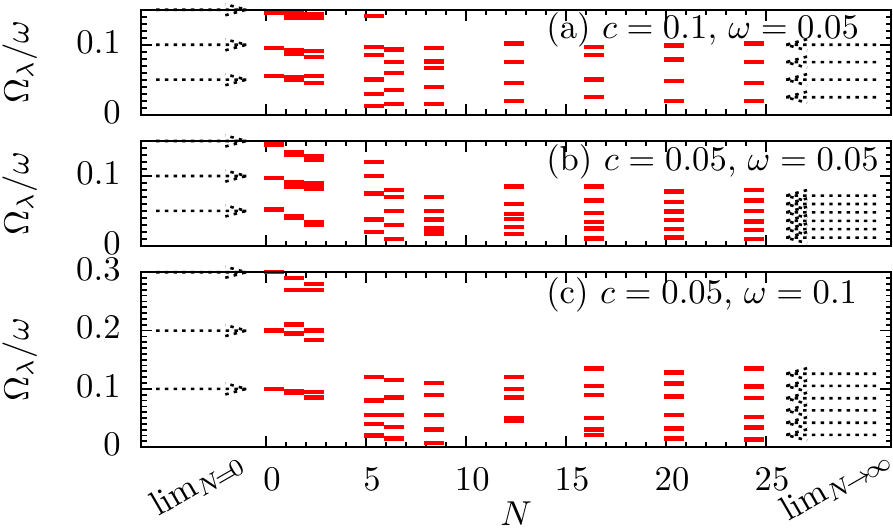}
 \caption{(Color online) The collective mode frequencies renormalized by the
   trapping potential $\Omega_{\lambda}/\omega$ as a function of number of
   embedded monopoles $N$ for different interaction strengths and trapping
   potentials. The dashed arrows highlight the expected modes in the $N=0$
   and $N\to\infty$ limits.}
 \label{fig:CollModes}
\end{figure}

On top of direct imaging, a second
experimental tool to characterize the BEC with embedded monopoles is the
collective mode spectrum. This was previously used to analyze both the
uniform and rotation-induced vortices in
BECs~\cite{Jin96,Edwards96,Stringari96,Dodd97,Coddington03}. Collective
modes can be excited by a red-detuned optical dipole potential that draws
atoms into the center of the condensate~\cite{Coddington03}.

To study the collective mode spectrum of the cloud we take the ground state
wavefunction and diagonalize the linear response Bogoliubov equations. In
\figref{fig:CollModes} we show the lowest mode frequencies for a series of
clouds with differing numbers $N$ of embedded monopoles. We first compare
the observed spectra at $N=0$ to that of a vortex-free BEC in a spherical
trap whose collective modes in the weakly interacting limit are simply the
single-particle modes in the parabolic trapping potential, so
$\Omega_{\lambda}=\lambda\omega$ with $\lambda\in\mathbb{Z}^{+}$. These
frequencies are in good agreement
with the calculated modes, and double in energy on doubling $\omega$ from
(b) to (c), but barely change on reducing the interaction strength $c$ from
(a) to (b). On increasing $N\lesssim6$ the flux quanta initially escape
along the $z$-axis, breaking the spherical symmetry, and splitting the
degenerate modes.  With a large number of embedded monopoles $N\to\infty$
the surface of the cloud resembles a two-dimensional BEC penetrated by a
vortex lattice (see for example \figref{fig:DensityPlots}(c)). The modes of
such a sheet are
$\Omega=h[nc(n_{\text{x}}^{2}+n_{\text{y}}^{2}+n_{\text{z}}^{2})/m]^{1/2}$~\cite{Sonin87},
with $\{n_{\text{x}},n_{\text{y}},n_{\text{z}}\}\in\mathbb{Z}$, $n$ the
density, and $L$ the sheet size, here the circumference of the cloud. These
modes were verified by their frequencies halving on reducing the interaction
strength from (a) to (b). This provides an intuitive picture of the low
energy collective modes in the $N\to\infty$ limit where the surface modes
dominate. With these signatures the collective modes could help diagnose the
behavior of the BEC embedded with monopoles.

In this paper we have constructed a gauge field of a line of magnetic
monopoles embedded in a BEC. The BEC underwent a transition from channeling
the flux along the Dirac string to the flux escaping out of the cloud walls.
Finally, we characterized this transition through the collective modes
spectrum. However, there are opportunities to explore monopolar gauge fields
in two other systems. The first is to study the single atom bound states in
a non-interacting gas. These feel the same effective magnetic field as the
condensate atoms, and could be characterized by exciting transitions between
different states. The second possibility is to generalize our single spin
species BEC up to a two-component BEC by exploiting a multi-level system
with doubly degenerate dark states and study
the emergence of topological states~\cite{Pietila09i}.

{\it Acknowledgments:} The author acknowledges the financial support of
Gonville \& Caius College.


\begin{thebibliography}{99}

\bibitem{Cabrera82}
B.~Cabrera, Phys. Rev. Lett., {\bf 48}, 1378 (1982).

\bibitem{Price75}
P.B.~Price, E.K.~Shirk, W.Z.~Osborne, and L.S.~Pinsky, Phys. Rev. Lett., {\bf 35}, 487 (1975). 

\bibitem{Castelnovo08}
C.~Castelnovo, R.~Moessner, and S.L.~Sondhi, Nature {\bf 451}, 42 (2008).

\bibitem{Qi09}
X.-L.~Qi {\it et al.}, Science {\bf 323}, 1184 (2009).

\bibitem{Fang03}
Z.~Fang {\it et al.}, Science {\bf 302}, 92 (2003).

\bibitem{Blaha76}
S.~Blaha, Phys. Rev. Lett., {\bf 36}, 874 (1976).

\bibitem{Dalibard11}
J.~Dalibard, F.~Gerbier, G.~Juzeli\=unas, and P.~\"Ohberg, Rev. Mod. Phys, {\bf 83},
1523 (2011).

\bibitem{Lin09}
Y.-J.~Lin, R.L.~Compton, K.~Jim\'enez-Garc\'ia, J.V.~Porto, and
I.B.~Spielman, Nature {\bf 462}, 628 (2009).

\bibitem{Lin08}
Y.-J.~Lin, R.L.~Compton, A.R.~Perry, W.D.~Phillips, J.V.~Porto, and I.B.~Spielman,
Phys. Rev. Lett. {\bf 102}, 130401 (2009).

\bibitem{Lin11i}
Y.-J.~Lin, K.~Jim\'enez-Garc\'ia, and I.B.~Spielman, Nature {\bf 471},
83 (2011).

\bibitem{Lin11ii}
Y.-J.~Lin, R.L.~Compton, K.~Jim\'enez-Garc\'ia, W.D.~Phillips, 
J.V.~Porto, and I.B.~Spielman, Nature Physics {\bf 7}, 531 (2011).

\bibitem{Cooper11}
N.R.~Cooper, Phys. Rev. Lett., {\bf 106}, 175301 (2011).

\bibitem{Aidelsburger11}
M.~Aidelsburger, M.~Atala, S.~Nascimb\`ene, S.~Trotzky, Y.-A.~Chen, and I.~Bloch, 
Phys. Rev. Lett. {\bf 107}, 255301 (2011).

\bibitem{Jaksch03}
D.~Jaksch and P.~Zoller, New J. Phys. {\bf 5}, 56 (2003).

\bibitem{Mueller04}
E.J.~Mueller, Phys. Rev. A {\bf 70}, 041603 (2004).

\bibitem{Gerbier10}
F.~Gerbier and J.~Dalibard, New J. Phys. {\bf 12}, 033007 (2010).

\bibitem{Sorensen05}
A.S.~S{\o}rensen, E.~Demler, and M.D.~Lukin,
Phys. Rev. Lett. {\bf 94}, 086803 (2005).

\bibitem{Alba11}
E.~Alba, X.~Fernandez-Gonzalvo, J.~Mur-Petit, J.K.~Pachos, and J.J.~Garcia-Ripoll,
Phys. Rev. Lett. {\bf 107}, 235301 (2011).

\bibitem{Moody86}
J.~Moody, A.~Shapere, and F.~Wilczek, \emph{Geometric Phases in Physics} (1989).

\bibitem{Ruseckas05}
J.~Ruseckas, G.~Juzeli\=unas, P.~\"Ohberg, and M.~Fleischhauer,
Phys. Rev. Lett., {\bf 95}, 010404 (2005).

\bibitem{Pietila09i}
V.~Pietil\"a, and M.~M\"ott\"onen, Phys. Rev. Lett., {\bf 102}, 080403 (2009).

\bibitem{Pietila09ii}
V.~Pietil\"a, and M.~M\"ott\"onen, Phys. Rev. Lett., {\bf 103}, 030401 (2009).

\bibitem{Wright00}
E.M.~Wright, J.~Arlt, and K.~Dholakia, Phys. Rev. A, {\bf 63}, 013608 (2000).

\bibitem{Olson07}
S.E.~Olson, M.L.~Terraciano, M.~Bashkansky, and F.K.~Fatemi, Phys. Rev. A, {\bf 76}, 061404(R) (2007).

\bibitem{John11}
S.T.~John, Z.~Hadzibabic, and N.R.~Cooper, Phys. Rev. A, {\bf 83}, 023610 (2011).

\bibitem{Spielman09}
I.B.~Spielman, Phys. Rev. A, {\bf 79}, 063613 (2009).

\bibitem{Edwards96}
M.~Edwards, P.A.~Ruprecht, K.~Burnett, R.J.~Dodd, and C.W.~Clark, Phys. Rev. Lett., {\bf 77}, 1671 (1996).

\bibitem{Dodd97}
R.J.~Dodd, K.~Burnett, M.~Edwards, and C.W.~Clark, Phys. Rev. A, {\bf 56}, 587 (1997).

\bibitem{Coddington03}
I.~Coddington, P.~Engels, V.~Schweikhard, and E.A.~Cornell, Phys. Rev. Lett., {\bf 91}, 100402 (2003).

\bibitem{Dalibard00}
K.W.~Madison, F.~Chevy, W.~Wohlleben, and J.~Dalibard, Phys. Rev. Lett. {\bf 84}, 806 (2000).

\bibitem{Engels02}
P.~Engels, I.~Coddington, P.C.~Haljan, V.~Schweikhard, and E.A.~Cornell,
Phys. Rev. Lett., {\bf 90}, 170405 (2003).

\bibitem{Shaeer01}
J.R.~Abo-Shaeer, C.~Raman, J.M.~Vogels, and W.~Ketterle, Science {\bf 292}, 476 (2001).

\bibitem{Stringari96}
S.~Stringari, Phys. Rev. Lett., {\bf 77}, 2360 (1996). 

\bibitem{Jin96}
D.S.~Jin, J.R.~Ensher, M.R.~Matthews, C.E.~Wieman, and E.A.~Cornell,
Phys. Rev. Lett., {\bf 77}, 420 (1996).

\bibitem{Sonin87}
E.B.~Sonin, Rev. Mod. Phys., {\bf 59}, 87 (1987).

\bibitem{Bruun01}
G.M.~Bruun, and L.~Viverit, Phys. Rev. A {\bf 64}, 063606 (2001).

\end{thebibliography}
\end{document}